\documentclass[twocolumn]{aastex631}
\usepackage{enumitem}

\begin{document}

\title{Evidence for a Fast Soft X-ray Wind in M82 from \textit{XMM}-RGS}

\correspondingauthor{Erin Boettcher}
\email{eboettch@umd.edu}

\author[0000-0003-3244-0409]{Erin Boettcher}
\affiliation{Department of Astronomy, University of Maryland, College Park, MD 20742, USA}
\affiliation{X-ray Astrophysics Laboratory, NASA/GSFC, Greenbelt, MD 20771, USA}
\affiliation{Center for Research and Exploration in Space Science and Technology, NASA/GSFC, Greenbelt, MD 20771, USA}

\author[0000-0002-2397-206X]{Edmund Hodges-Kluck}
\affiliation{X-ray Astrophysics Laboratory, NASA/GSFC, Greenbelt, MD 20771, USA}





\begin{abstract}

Starburst wind models predict that metals and energy are primarily carried out of the disk by hot gas ($T > 10^{6}$ K), but the low energy resolution of X-ray CCD observations results in large uncertainties on the mass and energy loading. Here, we present evidence for a fast soft X-ray wind from the prototypical starburst galaxy M82 using deep archival observations from the Reflection Grating Spectrometer on \textit{XMM-Newton}. After characterizing the complex line-spread function for the spatially extended outflow ($\approx 4'$), we perform emission-line fitting to measure the velocity dispersion, $\sigma_{\text{v}}$, from \ion{O}{8} (0.65, 0.77 keV), \ion{Ne}{10} (1.02 keV), and \ion{Mg}{12} (1.47 keV). For the $T \approx 3 \times 10^{6}$ K gas, \ion{O}{8} yields a velocity dispersion of $\sigma_{\text{v}} = 1160^{+100}_{-90}$ km s$^{-1}$, implying a wind speed that is significantly above the escape velocity ($v_{\text{esc}} \lesssim 450$ km s$^{-1}$). \ion{Ne}{10} ($\sigma_{\text{v}} = 550^{+130}_{-150}$ km s$^{-1}$) and \ion{Mg}{12} ($\sigma_{\text{v}} < 370$ km s$^{-1}$) show less velocity broadening than \ion{O}{8}, hinting at a lower wind speed or smaller opening angle on the more compact spatial scales traced by the $T \approx (0.6 - 1) \times 10^{7}$ K gas. Alternatively, these higher energy emission lines may be dominated by shock-heated gas in the interstellar medium. Future synthesis of these measurements with Performance Verification observations of the $E = 2 - 12$ keV wind in M82 from the Resolve microcalorimeter on the \textit{X-ray Imaging and Spectroscopy Mission} will inform the phase structure and energy budget of the hot starburst wind.

\end{abstract}



\section{Introduction} \label{sec:intro}

Starburst winds are the dominant feedback mechanism in low-mass galaxies. Driven by spatially and temporally clustered supernovae, they shape the star-formation efficiency in sub-$L^{*}$ galaxies and are responsible for redistributing mass, metals, and angular momentum from galaxy disks across circumgalactic and intergalactic scales (for a review, see \citealt{2005ARA&A..43..769V}). Cosmological models predict that star-formation driven winds with high mass-loading factors are increasingly common at high redshift, where the low-mass progenitors of $L^{*}$ galaxies experience vigorous episodes of star formation \citep[e.g.,][]{2015MNRAS.454.2691M}. This motivates careful studies of dwarf starburst winds in the local universe, where their multi-phase structural and kinematic properties can be probed at high spatial resolution (tens of parsecs).

A host of hydrodynamic simulations have been developed to explore the launching mechanisms, phase structure, and mass and energy loading of starburst winds \citep[e.g.,][]{2012MNRAS.421.3522H, 2016ApJ...821....7T, 2020ApJ...903L..34K, 2020ApJ...895...43S}. From these efforts, a general picture has emerged in which the cool ($T \lesssim 10^{4}$ K) and hot gas ($T > 10^{6}$ K) dominate the mass and energy loading, respectively. The hot phase is also believed to carry the majority of the metals injected by the ongoing star formation. However, while observational constraints on mass and energy loading in the $T \lesssim 10^{6}$ K gas have been enabled by techniques including down-the-barrel absorption-line spectroscopy in the rest-frame ultraviolet \citep[e.g.,][]{2014ApJ...794..156R, 2018MNRAS.474.1688C}, such constraints are largely lacking for the $T > 10^{6}$ K gas, which has been primarily probed with X-ray CCD spectroscopy. 

While X-ray CCD observations have provided a wealth of information about starburst winds, including their prevalence, temperature structure, and abundance patterns \citep[e.g.][]{2009ApJ...697.2030S, 2020ApJ...903...35H, 2020ApJ...904..152L, 2023ApJ...942..108L, 2023arXiv231208444P}, their typical energy resolution ($\Delta E \sim 100$ eV at $E = 1$ keV, or $\Delta v \sim 3 \times 10^{4}$ km s$^{-1}$) prevents precise measurements of the mass, metal, and energy loading in the $T \gtrsim 10^{6}$ K gas \citep[e.g.,][]{2000MNRAS.314..511S}. The total mass and metals carried by the wind is uncertain due to degeneracy between the density and metallicity in spectral modeling, and the total energy (thermal and kinetic) is unknown due to the unresolved velocity profiles of the hot gas. The temperature and chemical abundance determination is also affected by the possible presence of charge exchange (CX) emission, which is difficult to cleanly separate from thermal emission in CCD data. It is unclear whether the dominant, volume-filling phase of the outflow is generally in the soft ($T \sim 10^{6} - 10^{7}$ K) or hard ($T > 10^{7}$ K) X-ray regime, with the distinction determined by the mass loading and corresponding cooling efficiency \citep[e.g.,][]{2016MNRAS.455.1830T}; in the latter scenario, the soft X-rays may instead trace interfaces with cool gas, including CX emission. The velocity profile thus provides an important probe of the energy loading and an indirect diagnostic of the emission mechanism (e.g., whether the photons trace interfaces with cool clouds or a volume-filling phase of the wind).

By using a grating, dispersive X-ray spectroscopy can achieve an order of magnitude improvement in energy resolution compared to CCD observations, providing a direct measurement of the velocity profile of hot starburst winds. However, slitless, dispersive spectroscopy requires a relatively high X-ray surface brightness and produces a complex line-spread function (LSF) for spatially extended sources. Thus, this technique has yet to be applied to resolved starburst winds for the primary purpose of characterizing their kinematics. Here, we conduct the first comprehensive study of the velocity profile of a soft X-ray wind using archival observations of the dwarf starburst galaxy M82 from the Reflection Grating Spectrometer (RGS; \citealt{2001A&A...365L...7D}) on $XMM-Newton$.

As the prototypical dwarf starburst galaxy, M82 is a nearby target ($D = 3.6$ Mpc; $1' \approx 1$ kpc; \citealt{1994ApJ...427..628F}) whose multi-phase wind properties have been well characterized \citep[e.g.,][]{1998ApJ...493..129S, 2009ApJ...697.2030S, 2015ApJ...814...83L, 2018ApJ...856...61M, 2020ApJ...904..152L, 2023arXiv230915906L}. The outflow is driven by a central starburst (SFR $\sim 10$ M$_{\odot}$ yr$^{-1}$ within the central few hundred pc), with recent star-formation episodes peaking $\sim 5$ and $10$ Myr ago; this was likely tidally triggered by a close passage with its more massive companion, M81 \citep{2003ApJ...599..193F}. Due to its interaction history, the gaseous environment of M82 is complex, with evidence for tidal streams, a galactic fountain, and a fast outflow existing co-spatially \citep[e.g.,][]{2015ApJ...814...83L, 2023arXiv230915906L}.

The hot wind in M82 has been studied extensively over the past three decades \citep{1995ApJ...439..155B, 1997A&A...320..378S, 2000MNRAS.314..511S, 2003MNRAS.343L..47S, 2007ApJ...658..258S, 2009ApJ...697.2030S, 2011PASJ...63S.913K, 2020ApJ...904..152L}. \cite{2009ApJ...697.2030S} paired hydrodynamic simulations with hard X-ray line fluxes from the \textit{Chandra X-ray Observatory} and \textit{XMM-Newton} to infer a wind temperature of $T \approx 10^{7.5} - 10^{7.9}$ K within the starburst region, with a mass outflow rate of $\dot M_{\text{out}} \approx 1 - 4$ M$_{\odot}$ yr$^{-1}$. The predicted terminal velocity of the hard outflow is a factor of at least three to four above the escape velocity ($v_{\text{esc}} \lesssim 450$ km s$^{-1}$; \citealt{2009ApJ...697.2030S}), suggesting that the outflow will enrich the intragroup medium. Some simulations predict that the soft X-ray emitting gas is a sub-dominant component of starburst winds, occupying a small volume filling factor and carrying $\lesssim 10$\% of the mass and energy in the outflow \citep[e.g.,][]{2000MNRAS.314..511S}. The \textit{XMM}-RGS analysis presented here provides the first opportunity to directly test these predictions using the velocity profile of the soft X-ray wind in M82.

Previous studies have used \textit{XMM}-RGS observations of the M82 outflow for a variety of science goals. Early works by \citet{2002MNRAS.335L..36R} and \citet{2004ApJ...606..862O} found that the chemical abundance pattern in the hot gas is largely consistent with enrichment by Type II SNe, although discrepancy in the measured O abundances between the gas and stars suggested that unmodeled CX emission affects the inferred thermal and chemical properties of the gas \citep{2011MNRAS.415L..64L}. Later work reported significant CX emission based on the flux ratios of the K$\alpha$ triplets of He-like ions \citep{2008MNRAS.386.1464R, 2011MNRAS.415L..64L} and the broad-band spectrum \citep{2014ApJ...794...61Z, 2024A&A...686A..96F, 2024ApJ...963..147O}, implying a high surface area of interfaces between hot and cool gas. \citet{2024ApJ...963..147O} showed the possible presence of significant velocity broadening in the outflow by fitting velocity-broadened models to multiple archival spectra, yielding velocity dispersions that vary from $\sim 100$ km s$^{-1}$ to $\gtrsim 1000$ km s$^{-1}$ between observations. The variation in the best-fit velocity dispersion, as well as evidence for energy-dependent velocity broadening \citep[e.g.,][]{2002MNRAS.335L..36R}, motivates an analysis of the velocity profiles of individual emission lines across multiple, deep observations. We undertake that analysis here.

This paper is organized as follows. In \S~\ref{sec:data}, we present the archival data from \textit{XMM}-RGS and discuss the complexities of the RGS LSF for spatially extended sources. \S~\ref{sec:analysis} discusses the emission-line fitting used to measure the velocity dispersion from a small suite of soft X-ray lines. We present the measured velocity dispersions in \S~\ref{sec:results}, and we discuss their implications in \S~\ref{sec:disc}, including comparing the dispersions of the cool and hot winds and applying a simple biconical wind model to infer the wind velocity and opening angle. We summarize the implications for hot wind models in \S~\ref{sec:conc}. Throughout the paper, we assume a $\Lambda$ cosmology with $\Omega_{\text{M}} = 0.3$, $\Omega_{\Lambda} = 0.7$, and $H_{0} = 70$ km s$^{-1}$ Mpc$^{-1}$.

\section{Data} \label{sec:data}

\textit{XMM}-RGS provides slitless, dispersive spectroscopy in the soft X-ray band ($E = 0.3 - 2.1$ keV and $E = 0.65 - 3.0$ keV in first and second order, respectively). The field of view in the cross-dispersion direction is $4.4'$, and the full width at half maximum (FWHM) of the point-spread function (PSF) is $4'' - 5''$. The RGS consists of two identical modules, RGS1 and RGS2. 

We analyzed four archival observations of M82: the two deepest exposures (Obs ID: 0206080101; PI: P. Ranalli and Obs ID: 0560590301; PI: H. Feng, hereafter Obs 1 and Obs 2) and two shallower exposures taken at a position angle (PA) rotated approximately $180^{\circ}$ with respect to the previous observations (Obs ID: 0560590101; PI: H. Feng and Obs ID: 0891060101; PI: M. Bachetti, hereafter Obs 3 and Obs 4). The latter observations were analyzed to assess the impact of the PA-dependent LSF on the measurement of the intrinsic emission-line broadening in the spatially extended wind (see \S~\ref{sec:rmf}). A summary of the archival observations is given in Tab~\ref{tab:data}.

\begin{deluxetable*}{cccccccc}
  \tablecaption{Archival observations of M82 with $XMM$-RGS}
  \tablehead{
    \colhead{(1)} &
    \colhead{(2)} &
    \colhead{(3)} &
    \colhead{(4)} & 
    \colhead{(5)} &
    \colhead{(6)} &
    \colhead{(7)} &
    \colhead{(8)} \\
    \colhead{Obs. ID} &
    \colhead{PI} &
    \colhead{R.A. (J2000)} &
    \colhead{Decl. (J2000)} &
    \colhead{P.A.} & 
    \colhead{$t_{\text{exp}}$} &
    \colhead{$t_{\text{clean}}$} &
    \colhead{Obs. date} \\
    \colhead{} &
    \colhead{} &
    \colhead{(hh:mm:ss.ss)} &
    \colhead{(dd:mm:ss.s)} &
    \colhead{($^{\circ}$)} & 
    \colhead{(s)} &
    \colhead{(s)} &
    \colhead{MM/YY}
  }
  \startdata
  \noalign{\smallskip}
   0206080101 & P. Ranalli & 09:55:52.20 & +69:40:47.0 & 319 & 104353 & 80800 & 04/04  \\
   0560590301 & H. Feng & 09:55:50.19 & +69:40:47.0 & 296 & 52974 & 51200 & 04/09 \\
   0560590101 & H. Feng & 09:55:50.19 & +69:40:47.0 & 138 & 31914 & 31914 & 10/08 \\
   0891060101 & M. Bachetti & 09:55:51.03 & +69:40:45.5 & 125 & 29500 & 28500 & 10/21
  \smallskip
  \label{tab:data}
  \enddata
  \tablecomments{(1) Observation ID, (2) Program PI, (3) right ascension, (4) declination, (5) position angle, measured from North to East, (6) exposure time, (7) exposure time retained based on background level, and (8) date of observation.}
\end{deluxetable*}

We downloaded the data from the $XMM-Newton$ Science Archive\footnote{\url{https://nxsa.esac.esa.int/nxsa-web/\#home}}. Using the $XMM-Newton$ Scientific Analysis System\footnote{\url{https://www.cosmos.esa.int/web/xmm-newton/sas}} (SAS, v19.1.0), we reduced the RGS data using the \texttt{rgsproc} tool. To remove periods of high background, we excluded events that occurred when the count rate was $> 0.2$ cts/s in the background spectra recorded by CCD 9, which is most affected by proton flares. The retained exposure time was $\approx 80 - 100$\% of the total. We extracted the source spectra using apertures of width $2'$ in the cross-dispersion direction, centered on the pointing coordinates given in Tab.~\ref{tab:data} (see Fig.~\ref{fig:spatial_prof}). There are no clean regions on the RGS detectors from which to extract background spectra. However, as we expect the background to vary only slowly with energy, background counts do not affect our ability to measure the velocity broadening of emission lines, and we do not remove the background in our fitting. Our reduction yielded source spectra and point-source redistribution matrix files (RMFs) in first and second order for both of the RGS detectors.

\subsection{Response matrices}\label{sec:rmf}

\begin{figure*}
\centering
\includegraphics[scale = 0.55]{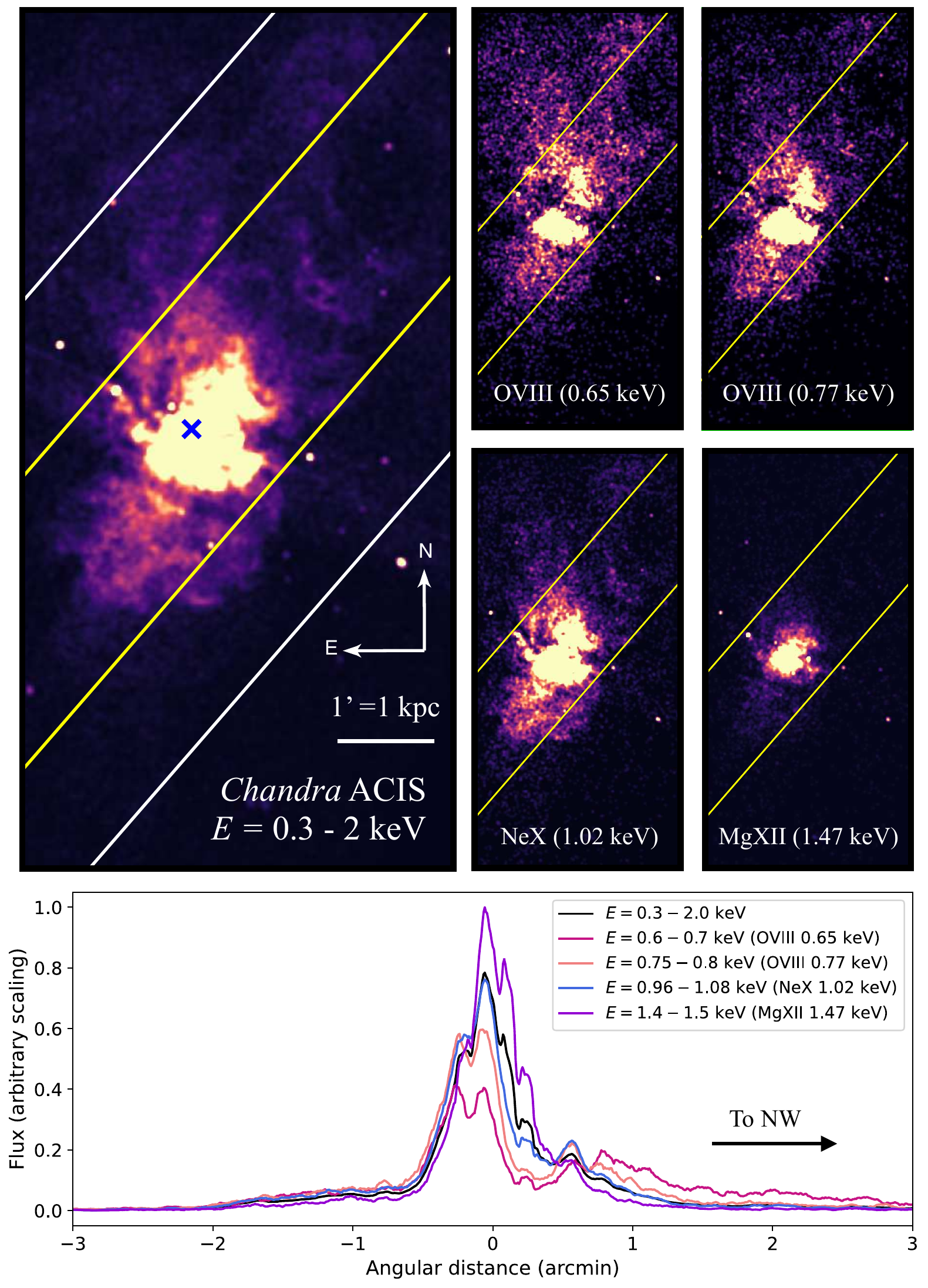}
\caption{An archival \textit{Chandra}-ACIS observation (Obs ID: 10542; PI: D. Strickland) shows that the soft X-ray emission from the M82 outflow ($E = 0.3 - 2$ keV) is spatially extended on scales of several arcminutes (top left panel). The spatial profile is significantly more extended and displays more substructure in the lowest energy emission lines (i.e., \ion{O}{8} 0.65 keV) than in the highest energy lines (\ion{Mg}{12} 1.47 keV; right and bottom panels). These differences are real and not attributable to counting statistics. Consequently, to measure the intrinsic emission-line broadening using the slitless RGS, the LSF must be convolved with a narrowband image at each emission line of interest. In the image panels, the white lines show the field of view of the RGS for Obs 1, centered at the blue cross; the yellow lines indicate the $2'$ source aperture. The image stretch is adjusted between panels for display purposes.}
\label{fig:spatial_prof}
\end{figure*}

The RGS is a slitless spectrometer, and consequently, the point-source LSF is convolved with the surface brightness profile at each wavelength for extended sources, which broadens and distorts the effective LSF. The pattern of the distortion in wavelength space can be calculated from the angular surface brightness distribution using one or more images of the soft X-ray emission. We used an archival data set from the Advanced CCD Imaging Spectrometer (ACIS) on \textit{Chandra} (Obs ID: 10542; PI: D. Strickland), which offers the highest spatial resolution view of the extended outflow ($\approx 1''$ FWHM on axis). The data were downloaded from the Chandra Data Archive\footnote{\url{https://cda.harvard.edu/chaser/}} and reduced using the \texttt{chandra\_repro} task in the Chandra Interactive Analysis of Observations\footnote{\url{https://cxc.cfa.harvard.edu/ciao/}} package (CIAO, v4.13).

\begin{figure}
\centering
\includegraphics[scale = 0.47]{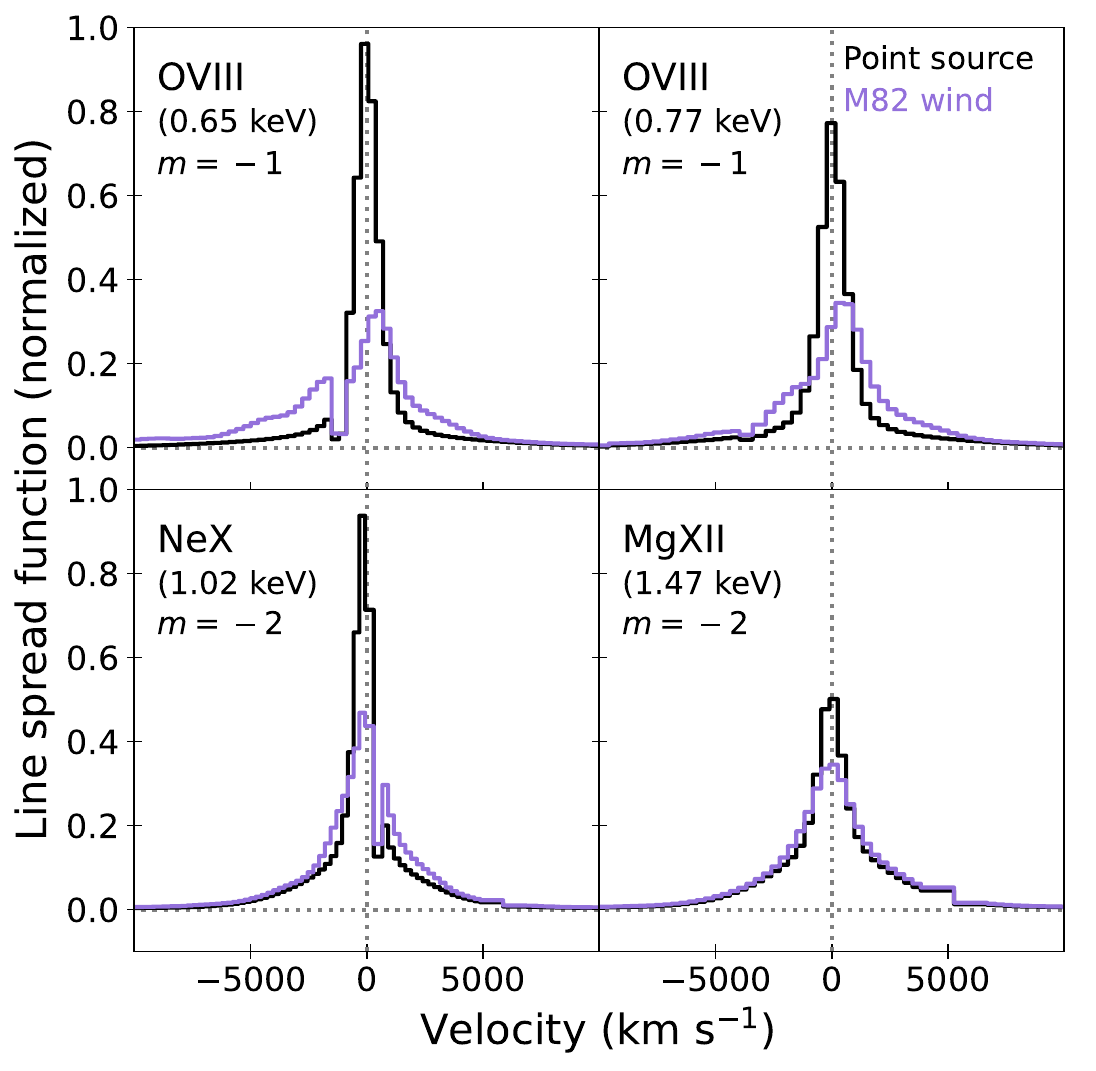}
\caption{The RGS LSF for key emission lines in first (top) and second order (bottom) for Obs 1. The LSF for point sources (black) is compared to the LSF for the spatially extended wind in M82 (purple), produced by convolving the point-source LSF with the spatial profile at the energy of the emission line of interest. There is significant instrumental line broadening due to the extended nature of the source, particularly on the blue wing of the \ion{O}{8} line.}
\label{fig:rmfs}
\end{figure}

The spatial profile is energy dependent, and this necessitates convolving the RGS LSFs with a narrowband image at each emission line of interest. As shown in Fig.~\ref{fig:spatial_prof}, we constructed a broadband, soft X-ray ($E = 0.3 - 2$ keV) image and narrowband images at the energies of each emission line. The narrowband images have energy widths of $\Delta E \approx 0.05 - 0.1$ keV, depending on the proximity of nearby emission features. The broadband emission is extended over $\gtrsim 4'$, and the spatial extent is greatest for the lowest energy emission lines (i.e., \ion{O}{8} 0.65 keV), which also show evidence for absorption in the disk plane. 

In Fig.~\ref{fig:rmfs}, we compare the LSFs for a point source with the LSFs for the M82 wind, produced by convolution with the \textit{Chandra} narrowband images. The convolution was performed using the \texttt{rgsxsrc} tool within the \texttt{Xspec}\footnote{\url{https://heasarc.gsfc.nasa.gov/xanadu/xspec/}} X-ray spectral fitting package (v12.12.1; see \S~\ref{sec:analysis}). We use high-resolution \textit{Chandra} images (on-axis PSF FWHM $\approx 1''$) instead of \textit{XMM}-MOS images because the PSF of the \textit{XMM} optics (on-axis FWHM $\approx 4''$) is already accounted for by \texttt{rgsxsrc} via the point-source RMF. Use of the MOS image thus artificially broadens the LSF by implementing the effect of the PSF twice (once through the point-source RMF and once through convolution with the MOS image). 

In first order, the full width at half maximum (FWHM) of the LSF for a point source is $\approx 1100$ km s$^{-1}$ at \ion{O}{8} ($0.65$ keV) and $2700$ km s$^{-1}$ at \ion{Mg}{12} ($1.47$ keV). For Obs 1, the FWHM becomes $\approx 3200$ km s$^{-1}$ and $4700$ km s$^{-1}$ at \ion{O}{8} and \ion{Mg}{12}, respectively, for the M82 wind. The second order spectrum, which has energy coverage blueward of \ion{O}{8} (0.65 keV), provides an improvement to FWHM $\approx 2800$ km s$^{-1}$ for the extended wind at \ion{Mg}{12}. There are significant wings on the LSFs for the extended sources, particularly on the \ion{O}{8} lines. 

As the PA of the observation changes, the dispersion direction rotates with respect to the asymmetric spatial profile of the wind. Obs 1 and 2 have similar PAs and therefore their convolved LSFs are comparable. Obs 3 and 4, however, are rotated by $181^{\circ}$ and $194^{\circ}$ with respect to Obs 1, respectively, and the extended LSFs for these exposures are comparable in width but inverted in velocity when compared to Obs 1 and 2 (i.e., the blue wing on \ion{O}{8} in Obs 1 becomes a red wing in Obs 3). This can help to separate intrinsic structure in the emission-line profiles from asymmetries induced by the LSFs. We apply these convolved responses in the following section to distinguish between intrinsic and instrumental emission-line broadening in the extended M82 outflow.

\section{Analysis} \label{sec:analysis}

\begin{figure*}
\centering
\includegraphics[scale = 0.45]{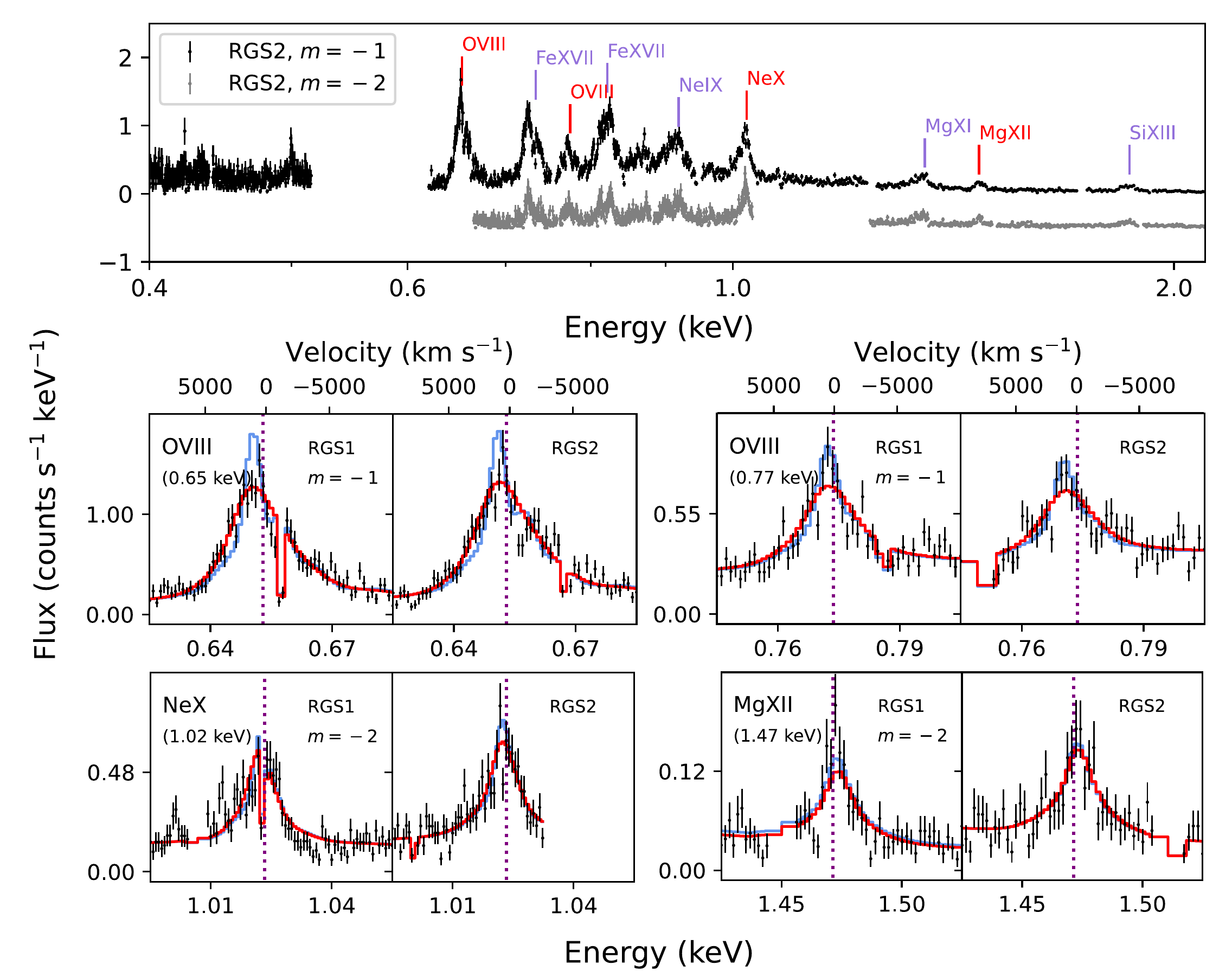}
\caption{Top panel: First- (black) and second-order (gray) RGS spectra of the M82 outflow from Obs 1. A flux offset is applied to the O2 spectrum for visual clarity. Emission lines labeled in red are included in our analysis. Bottom panels: Observed emission-line profiles (black points) and best-fit models (red curves) for \ion{O}{8} (O1; upper panels), \ion{Ne}{10} (O2; bottom left), and \ion{Mg}{12} (O2; bottom right). The models are shown for the best-fit $\sigma_{\text{v}}$ or $95\%$ upper limit for each line ($\sigma_{\text{v}}$ = 1160 km s$^{-1}$ for \ion{O}{8}, $\sigma_{\text{v}}$ = 550 km s$^{-1}$ for \ion{Ne}{10}, and $\sigma_{\text{v}}$ = 370 km s$^{-1}$ for \ion{Mg}{12}). The blue curves show the best-fit models with no velocity broadening ($\sigma_{\text{v}}$ = 0 km s$^{-1}$). The presence of velocity broadening is clear in the \ion{O}{8} 0.65 keV line profiles. The systemic redshift of M82 is indicated by the dotted purple lines. The \ion{O}{8} line profile peak is shifted away from the systemic velocity by the surface brightness distribution at these energies.}
\label{fig:spectra}
\end{figure*}

Our goal is to determine the velocity dispersion of the soft X-ray wind in M82 by measuring the intrinsic (non-instrumental) line broadening of a small suite of emission lines. We selected strong emission lines that optimize the following factors: 1) instrumental energy resolution, 2) lack of blending with nearby emission features, and 3) lack of charge exchange (CX) emission, which may be broadened differently from a more volume-filling component. We use the CX emission model for M82 from \citet{2014ApJ...794...61Z}. In first order (O1), these criteria are best satisfied by the \ion{O}{8} 0.65 keV and \ion{O}{8} 0.77 keV lines, which are relatively isolated from nearby emission features. However, the extended nature of the source produces a broad, asymmetric LSF at \ion{O}{8} (see Fig.~\ref{fig:rmfs}), and there may be a significant contribution from CX emission ($\approx (30 - 50)$\%; \citealt{2014ApJ...794...61Z, 2024ApJ...963..147O}). In second order (O2), the \ion{Ne}{10} 1.02 keV and \ion{Mg}{12} 1.47 keV lines provide the strongest constraints; they are relatively isolated and likely to have only minor contributions from CX. The full first- and second-order RGS spectra are shown in Fig.~\ref{fig:spectra}.

As the peak gas temperature probed evolves with emission-line energy ($T \approx 3 \times 10^{6}$ K and $T \approx 10^{7}$ K at \ion{O}{8} and \ion{Mg}{12}, respectively), we measure the emission-line broadening for each line separately. This yields four measurements, derived from \ion{O}{8} 0.65 keV (O1), \ion{O}{8} 0.77 keV (O1), \ion{Ne}{10} (O2), and \ion{Mg}{12} (O2). We additionally obtain a fifth measurement by fitting the \ion{O}{8} lines simultaneously. For all measurements, we fit jointly between the RGS1 and RGS2 detectors. For the \ion{O}{8} lines, we neglect the second order spectra because they lack coverage of the 0.65 keV line and have insufficient signal-to-noise ratio (SNR) at the 0.77 keV line to improve the constraint. Similarly, for the \ion{Ne}{10} and \ion{Mg}{12} lines, we do not include the first order spectra because the energy resolution is insufficient to improve the measurement (RGS1 also lacks first-order coverage of \ion{Ne}{10}, due to a failed CCD).

For each measurement, we fit emission-line models using the \texttt{rgsxsrc} tool within \texttt{Xspec} to account for the energy-dependent, spatially extended response (see \S~\ref{sec:rmf}). The models consist of a single Gaussian function and a powerlaw distribution to account for any local continuum:
\begin{equation}
f(E) = K_{0}[E(1 + z)]^{-\alpha} + \frac{K_{1}}{\sqrt{2\pi}\sigma_{\text{E}}(1 + z)}\mathrm{e}^{-(E(1 + z) - E_{0})^{2}/2\sigma_{\text{E}}^{2}}.
\end{equation}
Here, $f(E)$ is the model flux as a function of energy, $E$. The model has five free parameters: the redshift, $z$, the powerlaw normalization, $K_{0}$, the powerlaw index, $\alpha$, the Gaussian normalization, $K_{1}$, and the standard deviation of the line in the rest frame in keV, $\sigma_{\text{E}}$. $E_{0}$ denotes the line energy in the rest frame. We fit a single Gaussian model rather than an \texttt{APEC} model \citep{2001ApJ...556L..91S} for several reasons. First, the temperature structure of the outflow is complex, and multiple temperature components are likely present \citep[e.g.,][]{2024ApJ...963..147O}. Second, CX may contribute significantly to the wind emission, but there is not consensus on the CX model and its fractional contribution to the total flux in important diagnostic emission lines \citep[e.g.,][]{2014ApJ...794...61Z, 2020ApJ...904..152L, 2024A&A...686A..96F, 2024ApJ...963..147O}. We do not attempt to disentangle these complexities here, and instead aim to characterize the emission-line profiles independent of the choice of a thermal or CX model. We note, however, that we do not find significant differences when fitting single \texttt{APEC} models as when fitting Gaussian profiles. We focus on quantifying the emission-line widths rather than their centroids because of uncertainty in the RGS wavelength scale due to the sun angle correction.

Adopting the single Gaussian model, we use the \texttt{chain} implementation of the Goodman-Weare Markov Chain Monte Carlo (MCMC) algorithm in \texttt{Xspec} to perform the fit. The MCMC chains have a length of $N = 10^{5}$ steps and consist of $50$ walkers. For each emission line, this yields marginalized probability density functions (PDFs) for the velocity dispersion, $\sigma_{\text{v}}$, which we discuss in \S~\ref{sec:results}.

\section{Results} \label{sec:results}

\begin{figure}
\centering
\includegraphics[scale = 0.35]{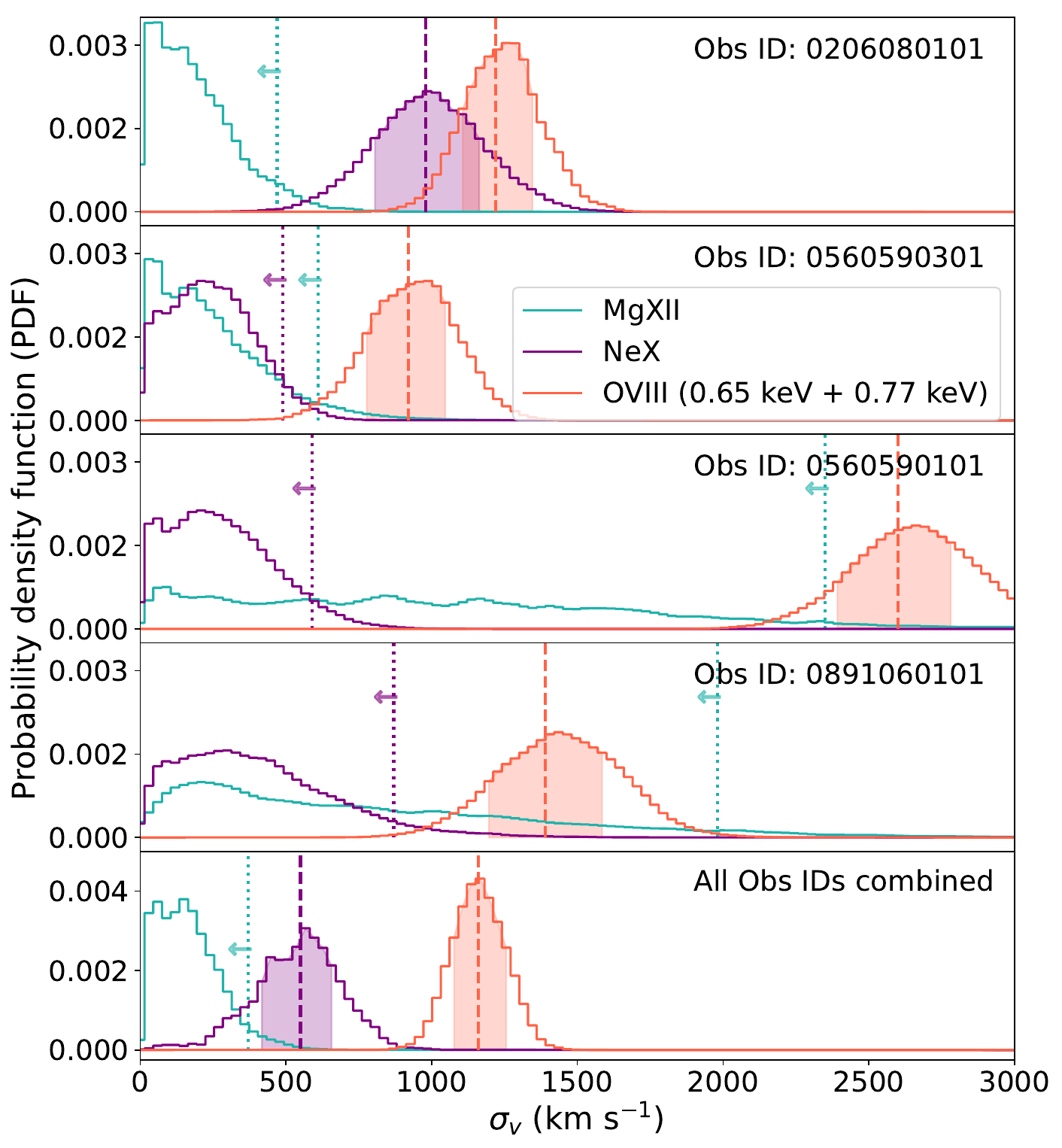}
\caption{The velocity dispersion, $\sigma_{\text{v}}$, of the soft X-ray wind from M82 is broad at \ion{O}{8} ($\sigma_{\text{v}} = 1160^{+100}_{-90}$ km s$^{-1}$) and is narrower in the hotter, more spatially compact gas traced by \ion{Ne}{10} ($\sigma_{\text{v}} = 550^{+130}_{-150}$ km s$^{-1}$) and \ion{Mg}{12} ($\sigma_{\text{v}} \lesssim 370$ km s$^{-1}$). PDFs for $\sigma_{\text{v}}$ measured from \ion{Mg}{12}, \ion{Ne}{10}, and \ion{O}{8} are indicated in cyan, purple, and red, respectively. The top four panels show measurements from Obs 1 - Obs 4; Obs 1, 2, and 4 are combined via a weighted product ($w_{i} = \sqrt{t_{\text{clean,}i}}$) in the bottom panel (see the text for a discussion of Obs 3). The dotted lines indicate 95\% upper limits, the dashed lines mark $50^{\text{th}}$ percentile values, and the shaded regions indicate the 68\% confidence intervals.}
\label{fig:pdfs}
\end{figure}

Here we report measurements of the velocity dispersion, $\sigma_{\text{v}}$, of the $T \approx (0.3 - 1) \times 10^{7}$ K outflow from M82. We find a best-fit value of $\sigma_{\text{v}} = 1160^{+100}_{-90}$ km s$^{-1}$ for \ion{O}{8} (0.65 keV and 0.77 keV) and $\sigma_{\text{v}} = 550^{+130}_{-150}$ km s$^{-1}$ for \ion{Ne}{10} (1.02 keV) and a $95$\% upper limit of $\sigma_{\text{v}} < 350$ km s$^{-1}$ for \ion{Mg}{12} (1.47 keV). We report $\sigma_{\text{v}}$ in Tab.~\ref{tab:sigma}, and the corresponding PDFs are shown in Fig.~\ref{fig:pdfs}. Example best-fit emission-line profiles are compared to the data in the bottom panels of Fig.~\ref{fig:spectra}. Below, we discuss the results from \ion{O}{8} for the $T \approx 3 \times 10^{6}$ K gas in \S~\ref{sec:results_OVIII} and from \ion{Ne}{10} and \ion{Mg}{12} for the $T \approx (0.6 - 1) \times 10^{7}$ K gas in \S~\ref{sec:results_NeX} and \S~\ref{sec:results_MgXII}, respectively.

\subsection{\ion{O}{8} results} \label{sec:results_OVIII}

The \ion{O}{8} ions trace the coolest ($T \approx 3 \times 10^{6}$ K) and most spatially extended ($l \gtrsim 4$ kpc) gas accessible with the RGS (see Fig.~\ref{fig:spatial_prof}). We first consider the results from Obs 1, the deepest exposure available. As shown in Fig.~\ref{fig:pdfs}, a joint fit to the \ion{O}{8} 0.65 keV and 0.77 keV lines yields a broad velocity distribution ($\sigma_{\text{v}} = 1220 \pm 130$ km s$^{-1}$). We next compare with the results from Obs 2, which were taken at a similar PA and serve as a consistency check. The best-fit velocity dispersion at \ion{O}{8}, $\sigma_{\text{v}} = 920^{+150}_{-160}$ km s$^{-1}$, is consistent with Obs 1 within $1.5\sigma$. We last consider the measurements from Obs 3 and Obs 4, which are taken at PAs that are $\sim 180^{\circ}$ offset from Obs 1 and Obs 2, inverting the LSFs in velocity between the two observations. This provides leverage on the extent to which the model fitting appropriately distinguishes between intrinsic and instrumental line broadening. Indeed, we find that the \ion{O}{8} velocity dispersion measured in Obs 4, $\sigma_{\text{v}} = 1390 \pm 210$ km s$^{-1}$, is consistent with Obs 1 and Obs 2 within $< 1\sigma$ and $2\sigma$, respectively, suggesting that the instrumental line broadening is appropriately modeled.

In contrast, Obs 3 yields a much larger best-fit value of $\sigma_{\text{v}} = 2600 \pm 210$ km s$^{-1}$, which is inconsistent with the other three observations at high significance. We discuss this observation in more detail in Appendix~\ref{sec:ulx}. In summary, both the \ion{O}{8} 0.65 keV line and the \ion{O}{7} 0.57 keV triplet show a secondary peak that is inconsistent with any LSF or instrumental effect. This feature is real, astrophysical, and present in both RGS modules. We discuss the possibility that this feature arises from a wind associated with one of the ultraluminous X-ray sources (ULXs) in M82 in Appendix~\ref{sec:ulx}. We report the fitting results from this observation for completeness, but we exclude this data set when constructing joint PDFs for $\sigma_{\text{v}}$ across the observations.

As a further consistency check on the measured \ion{O}{8} velocity dispersion, we fit for the dispersion separately from the \ion{O}{8} 0.65 keV and 0.77 keV lines in all observations. As seen in Tab.~\ref{tab:sigma}, the lower SNR in the weaker 0.77 keV transition results largely in upper limits, which are generally consistent with the 0.65 keV measurements. However, these are in tension for Obs 1, where the best-fit value from the 0.65 keV line falls almost $3\sigma$ above the $95\%$ upper limit from the 0.77 keV line. In addition to its lower SNR, the \ion{O}{8} 0.77 keV line is also blended with an \ion{Fe}{18} transition, both of which may affect the fidelity of the measurement. The most reliable measurement of $\sigma_{\text{v}}$ comes from the \ion{O}{8} 0.65 keV line, which optimizes SNR and energy resolution and is relatively isolated, albeit likely affected by CX emission. We report the \ion{O}{8} 0.77 keV measurements for completeness, and we weight the 0.65 keV and 0.77 keV lines by their SNRs when determining the joint \ion{O}{8} (0.65 keV + 0.77 keV) fit reported in Tab.~\ref{tab:sigma}.

In the bottom panel of Fig.~\ref{fig:pdfs}, we construct final, joint PDFs of the velocity dispersion by taking a weighted product of the PDFs from the individual observations (Obs 1, 2, and 4). We weight the PDFs by a proxy for the SNR, $w_{i} = \sqrt{t_{\text{clean,}i}}$, where $t_{\text{clean,}i}$ is the clean exposure time in the $i$\textsuperscript{th} observation after excluding time intervals of high background. This yields a final, best-fit value of $\sigma_{\text{v}} = 1160^{+100}_{-90}$ km s$^{-1}$ for the $T \approx 3 \times 10^{6}$ K gas traced by \ion{O}{8}.

We verified that the large measured velocity dispersion for \ion{O}{8} is not caused by the complex LSF for the extended \ion{O}{8}-bearing wind as follows. There are two choices that are made when using \texttt{rgsxsrc} to construct the extended source LSF: 1) the source image (e.g., \textit{Chandra}-ACIS vs. \textit{XMM}-MOS), and 2) the extraction aperture. In literature analyses of M82 RGS observations, a variety of choices have been made for both parameters \citep[e.g.,][]{2008MNRAS.386.1464R, 2014ApJ...794...61Z, 2024A&A...686A..96F, 2024ApJ...963..147O}. As discussed in \S~\ref{sec:rmf}, we use \textit{Chandra}-ACIS images, the highest angular resolution images available, to avoid double counting the impact of the \textit{XMM} PSF, which is already accounted for in the point-source LSF.

For the extraction aperture, we use the same aperture in \texttt{rgsxsrc} as we use to extract the source spectra ($2'$). As the \ion{O}{8} emission extends beyond the source extraction aperture (see Fig.~\ref{fig:spatial_prof}), the extended source LSFs in principle neglect a small amount of flux that is scattered into the aperture by the PSF wings. An examination of the observed \ion{O}{8} 0.65 keV line profiles and extended LSFs indicates that the line core dominates the measured broadening, suggesting scattered light from outside the extraction aperture should have a minimal impact. To quantify this, we considered two limiting cases for the \texttt{rgsxsrc} extraction aperture: 1) $2'$, equal to the source extraction aperture and the case reported here, and 2) $4'$, comparable to the RGS FOV. The former case will slightly overestimate the broadening in the line wings, while the latter case will underestimate the broadening, as not all emission outside of the source extraction aperture is scattered into the aperture by the PSF. The best-fit velocity dispersion for these cases should bracket the true dispersion, with the true solution likely closer to the former case due to the centrally peaked surface brightness profile. A joint fit to Obs 1, 2, and 4 conducted with an aperture of $4'$ ($2'$) in \texttt{rgsxsrc} yields a velocity dispersion of $\sigma_{\text{v}} = 850^{+110}_{-90}$ km s$^{-1}$ ($\sigma_{\text{v}} = 1210 \pm 100$ km s$^{-1}$) for \ion{O}{8} 0.65 keV. Thus, while the \texttt{rgsxsrc} aperture introduces modest systematic uncertainty in the \ion{O}{8} line broadening, there is significant instrinsic velocity broadening in the \ion{O}{8}-bearing wind ($\sigma_{\text{v}} \gtrsim 900$ km s$^{-1}$). We report the best-fit velocity dispersions determined for an aperture of $2'$ for the remainder of this work.

\begin{deluxetable*}{cccccc}
  \tablecaption{Velocity dispersion, $\sigma_{\text{v}}$, in the soft X-ray wind from M82}
  \tablehead{
    \colhead{(1)} &
    \colhead{(2)} &
    \colhead{(3)} &
    \colhead{(4)} &
    \colhead{(5)} & 
    \colhead{(6)} \\
    \colhead{Emission line} &
    \colhead{$\sigma_{\text{v}}$ (km s$^{-1}$)} &
    \colhead{$\sigma_{\text{v}}$ (km s$^{-1}$)} &
    \colhead{$\sigma_{\text{v}}$ (km s$^{-1}$)} &
    \colhead{$\sigma_{\text{v}}$ (km s$^{-1}$)} &
    \colhead{$\sigma_{\text{v}}$ (km s$^{-1}$)} \\
    \colhead{} &
    \colhead{(0206080101)} &
    \colhead{(0560590301)} &
    \colhead{(0560590101)} &
    \colhead{(0891060101)} &
    \colhead{(Combined)}
  }
  \startdata
  \noalign{\smallskip}
   \ion{O}{8} (0.65 keV) & $1310^{+150}_{-140}$ & $960^{+160}_{-170}$ & $2670 \pm 210$ & $1410^{+220}_{-230}$ & $1210 \pm 100$ \\
   \ion{O}{8} (0.77 keV) & $< 930$ & $< 1173$ & $< 3330$ & $1500^{+430}_{-380}$ & $740^{+190}_{-210}$ \\
   \ion{O}{8} (0.65 keV + 0.77 keV) & $1220 \pm 130$ & $920^{+150}_{-160}$ & $2600 \pm 210$ & $1390 \pm 210$ & $1160^{+100}_{-90}$ \\
   \ion{Ne}{10} (1.02 keV) & $980 \pm 190$ & $< 490$ & $< 590$ & $< 870$ & $550^{+130}_{-150}$ \\
   \ion{Mg}{12} (1.47 keV) & $< 470$ & $< 610$ & $< 2350$ & $< 1980$ & $< 370$
   \smallskip
  \label{tab:obs}
  \enddata
  \tablecomments{(1) Emission line(s) and (2) - (5) best-fit velocity dispersions, $\sigma_{\text{v}}$, from Obs 1 - 4. In column (6), we report a joint measurement using Obs 1, 2, and 4 (see text for discussion of Obs 3). Best-fit values are the $50$\textsuperscript{th} percentile values, error bars are the $68$\% confidence intervals, and limits are $95$\% upper limits.}
  \label{tab:sigma}
\end{deluxetable*}

\subsection{\ion{Ne}{10} results} 
\label{sec:results_NeX}

\ion{Ne}{10} 1.02 keV does not provide as clean of a diagnostic as \ion{O}{8} 0.65 keV. It is blended with an \ion{Fe}{17} transition, and while some models predict that the CX contribution is relatively low ($\approx 10$\%; \citealt{2014ApJ...794...61Z}), the Ly$\beta$/Ly$\alpha$ line ratio suggests that CX may contribute significantly to the \ion{Ne}{10} emission near the disk \citep{2016MNRAS.458.3554C}. Nevertheless, we consider \ion{Ne}{10} as a tracer of velocity broadening in the phase of the outflow with an intermediate temperature ($T \approx 6 \times 10^{6}$ K) and spatial extent (see Fig.~\ref{fig:spatial_prof}) compared to \ion{O}{8} and \ion{Mg}{12}. 

Beginning with Obs 1, the \ion{Ne}{10} velocity dispersion is $\sigma_{\text{v}} = 980 \pm 190$ km s$^{-1}$, consistent with the best-fit value from \ion{O}{8} within $1\sigma$. The remaining three observations yield upper limits that range from $\sigma_{\text{v}} < 490$ km s$^{-1}$ (Obs 2) to $\sigma_{\text{v}} < 870$ km s$^{-1}$ (Obs 4). Obs 1 is formally consistent with Obs 2 - 4 within $\lesssim 3\sigma$, but there is tension between the measurements. The joint PDF constructed from Obs 1, 2, and 4 yields a best-fit value of $\sigma_{\text{v}} = 550^{+130}_{-150}$ km s$^{-1}$. Thus, while it is likely that the \ion{Ne}{10} line is at least modestly narrower than \ion{O}{8}, additional deep observations are needed to definitely determine the velocity broadening of \ion{Ne}{10}.

\subsection{\ion{Mg}{12} results} 
\label{sec:results_MgXII}

\ion{Mg}{12} 1.47 keV is isolated from other emission lines and likely has only a minor contribution from CX emission ($\approx 10$\%; \citealt{2014ApJ...794...61Z}), making it a robust tracer of the velocity broadening in the hottest ($T \approx 10^{7}$ K) and most spatially compact ($l \sim 1$ kpc; see Fig.~\ref{fig:spatial_prof}) gas observed by the RGS. In Obs 1, the velocity dispersion PDF rises monotonically as $\sigma_{\text{v}}$ approaches $\sigma_{\text{v}} = 0$ km s$^{-1}$, with an upper limit at $\sigma_{\text{v}} < 470$ km s$^{-1}$. As shown in Tab.~\ref{tab:sigma}, the upper limits measured in Obs 2 - 4 become increasingly less constraining as the SNR declines. The upper limit determined jointly from Obs 1, 2, and 4 is $\sigma_{\text{v}} < 370$ km s$^{-1}$, demonstrating that the $T \approx 10^{7}$ K gas traced by \ion{Mg}{12} has considerably less velocity broadening than the $T \approx 3 \times 10^{6}$ K gas traced by \ion{O}{8}. In the following discussion, we focus on interpreting the \ion{O}{8} and \ion{Mg}{12} line widths in the context of the structure and kinematics of the multi-phase outflow.

\section{Discussion} \label{sec:disc}

The velocity dispersion and spatial extent of the coolest ($T \approx 3 \times 10^{6}$ K; \ion{O}{8}-bearing) and hottest ($T \approx 10^{7}$ K; \ion{Mg}{12}-bearing) gas traced by RGS, in combination with velocity measurements of the cool ($T \lesssim 10^{4}$ K) wind from the literature, constrains (a) the emission mechanism of the soft X-ray photons, and (b) the wind velocity and opening angle. We first compare the velocity dispersion of the soft X-ray wind with that of the cool gas traced by CO, \ion{H}{1} 21-cm, and H$\alpha$ emission to assess the likelihood that the soft X-ray photons arise primarily at interfaces in \S~\ref{sec:cool_wind}. We then consider the velocity and opening angle of the hot wind in \S~\ref{sec:wind_struct}.

\subsection{Comparison with the cool wind} \label{sec:cool_wind}

In this section, we demonstrate that the \ion{O}{8}-bearing outflow is unlikely to arise exclusively at the boundaries between a hotter wind fluid and cool, entrained clouds (i.e., produced by rapidly cooling gas or CX emission at interfaces). If \ion{O}{8} is primarily an interface species, then we would expect the kinematics of the cool and \ion{O}{8}-bearing gas to be comparable. However, as discussed below, the dispersion of the cool gas is several times smaller than that of the $T \approx 3 \times 10^{6}$ K gas, suggesting that the \ion{O}{8} may arise from a more volume-filling phase of the outflow.

The M82 outflow is highly multiphase, with molecular \citep{2015ApJ...814...83L}, neutral \citep{2018ApJ...856...61M}, and warm ionized gas \citep{1998ApJ...493..129S, 2007PASP..119....1M} as well as dust and polycyclic aromatic hydrocarbons \citep{2010A&A...518L..66R, 2010A&A...514A..14K} detected within and around the hot outflow cone. The $T \lesssim 10^{4}$ K gas shows evidence of a double-peaked velocity profile \citep[e.g.,][]{1998ApJ...493..129S, 2015ApJ...814...83L}, suggesting that it is found primarily at the edges of the cone. However, multiple physical processes likely contribute to producing the cool extraplanar gas, including acceleration of cold clouds via ram pressure in the hot wind \citep[e.g.,][]{2022MNRAS.510..551F}, ejective star-formation feedback that drives a cold fountain flow \citep{2018ApJ...856...61M}, and tidal stripping that creates cool streams and debris in the M81 group environment \citep[e.g.,][]{2023arXiv230915906L}.

CO (2-1) and \ion{H}{1} 21-cm emission are both detected within $\Delta v \approx \pm 150$ km s$^{-1}$ of the systemic velocity \citep{2018ApJ...856...61M, 2015ApJ...814...83L}.  In the warm ionized phase, the H$\alpha$ centroid of the primary velocity component is spread over $\Delta v \approx \pm 175$ km s$^{-1}$, with a secondary, high-velocity component observed over $\Delta v \approx \pm 400$ km s$^{-1}$ \citep{1998ApJ...493..129S}. Using IFU spectroscopy, \citet{2009ApJ...696..192W} further decompose the nebular line profiles in the inner wind into multiple velocity components that are generally separated by $\Delta v \lesssim 150$ km s$^{-1}$, with the broadest components having a typical $\sigma_{\text{v}} \approx 100$ km s$^{-1}$. Similarly, \citet{2023arXiv230915906L} report a characteristic velocity dispersion of $\sigma_{\text{v}} = 100$ km s$^{-1}$ traced by [\ion{C}{2}] emission in the southern outflow cone. In summary, the molecular, neutral, and warm ionized gas yield a generally consistent picture in which the flux-weighted, spatially integrated velocity dispersion of the $T \lesssim 10^{4}$ K gas does not exceed several hundred km s$^{-1}$.

The velocity dispersion of the $T \approx 3 \times 10^{6}$ K gas traced by \ion{O}{8}, $\sigma_{\text{v}} = 1160^{+100}_{-90}$ km s$^{-1}$, significantly exceeds that of the $T \lesssim 10^{4}$ K gas, falling $> 5\sigma$ above its characteristic dispersion. While this does not rule out a contribution of interface emission to the \ion{O}{8} signal, it suggests that interfaces are not the primary origin of the $T \approx 3 \times 10^{6}$ K gas. Possible alternative origins include adiabatically cooling wind fluid and gas that has been shock heated by a hotter, faster wind. For the $T \approx 10^{7}$ K gas traced by \ion{Mg}{12}, the upper limit on the velocity dispersion ($\sigma_{\text{v}} \lesssim 370$ km s$^{-1}$) cannot rule out an interface origin. In the following section, we further discuss the relationship between the \ion{O}{8}- and \ion{Mg}{12}-bearing gas in the context of a simple, biconical wind model.

\subsection{Geometry and speed of the soft X-ray wind} \label{sec:wind_struct}

\begin{figure*}
\centering
\includegraphics[scale = 0.28]{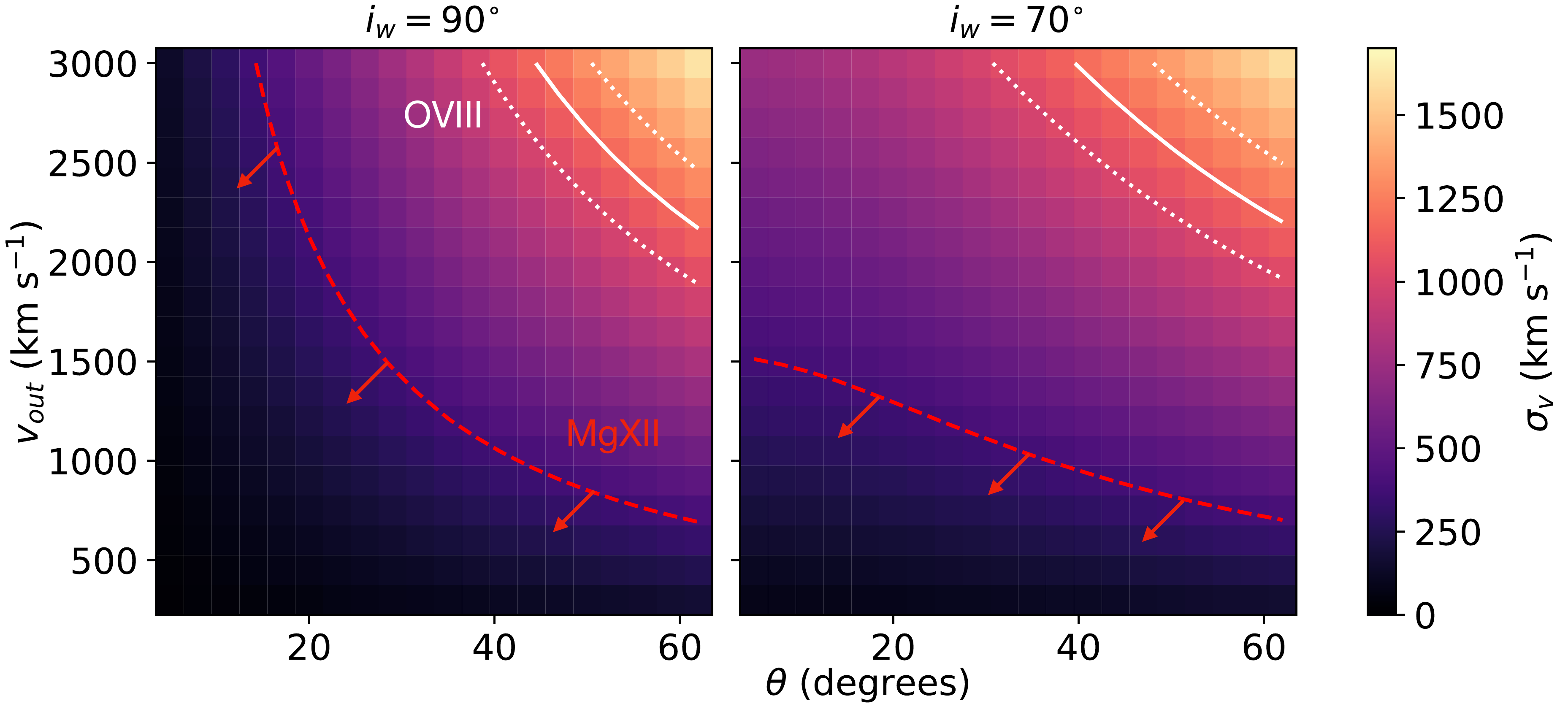}
\caption{The \ion{O}{8} velocity dispersion suggests a fast soft X-ray wind in M82. The expected velocity dispersion of a biconical outflow viewed edge-on ($i_{\text{w}} = 90^{\circ}$, at left) and at an inclination ($i_{\text{w}} = 70^{\circ}$, at right) is shown as a function of the radial outflow speed, $v_{\text{out}}$, and opening angle, $\theta$. This is compared to the best-fit value (white solid line) and $90$\% confidence interval (white dotted lines) measured for the $T \approx 3 \times 10^{6}$ K gas traced by \ion{O}{8}. The \ion{O}{8} line width suggests a fast outflow speed ($v_{\text{out}} \gtrsim 2000$ km s$^{-1}$) for both inclinations. The $95$\% upper limit measured from \ion{Mg}{12}, shown by the dashed red line, can accommodate a lower $v_{\text{out}}$ or a smaller $\theta$ on the more compact spatial scales traced by the $T \approx 10^{7}$ K gas.}
\label{fig:wind_cone_model}
\end{figure*}

\begin{deluxetable}{ccc}
  \tablecaption{\ion{O}{8} wind speed predicted by biconical wind model}
  \tablehead{
    \colhead{} &
    \colhead{$v_{\text{out}}$ (km s$^{-1}$)} &
    \colhead{} \\
    \colhead{} &
    \colhead{$i_{\text{w}} = 90^{\circ}$} &
    \colhead{$i_{\text{w}} = 70^{\circ}$}
  }
  \startdata
  \noalign{\smallskip}
  $\theta = 20^{\circ}$ & 6700 & 4050 \\
  $\theta = 40^{\circ}$ & 3350 & 3000 \\
  $\theta = 60^{\circ}$ & 2250 & 2250
  \smallskip
  \label{tab:wind_model}
  \enddata
  \tablecomments{Radial wind speed required to reproduce observed \ion{O}{8} velocity dispersion, $\sigma_{\text{v}} = 1160^{+100}_{-90}$ km s$^{-1}$, in a biconical wind with inclination angle $i_{\text{w}}$ and opening angle $\theta$.}
\end{deluxetable}

Here, we demonstrate that the velocity dispersion of the $T \approx 3 \times 10^{6}$ K gas, $\sigma_{\text{v}} = 1160^{+100}_{-90}$ km s$^{-1}$, implies a fast radial wind speed ($v_{\text{out}} \gtrsim 2000$ km s$^{-1}$) and large energy loading factor ($E_{\text{kin}}/E_{\text{th}} \gtrsim 25$), consistent with expectations from hot wind models. The much smaller velocity dispersion of the $T \approx 10^{7}$ K phase, $\sigma_{\text{v}} < 370$ km s$^{-1}$, suggests a different origin for this gas, such as shocked interstellar medium (ISM) surrounding the circumnuclear star clusters. Alternatively, this more spatially compact gas may arise from the base of the outflow if the wind speed or opening angle increases with distance from the disk.

We consider a simple model for a volume-filling, biconical outflow with a constant radial wind speed, $v_{\text{out}}$, and a fixed opening angle, $\theta$. The opening angle is defined with respect to the minor axis (i.e., the full angular extent of the cone is $2\theta$). We then construct the spatially integrated velocity profile, assigning equal luminosity weighting to equal volume elements. In Fig.~\ref{fig:wind_cone_model}, we show the modeled velocity dispersion (approximated as the standard deviation of the velocity distribution) as a function of $v_{\text{out}}$ and $\theta$. As the inclination angle of the hot wind, $i_{\text{w}}$, may not align with the inclination of the galaxy ($i \approx 81^{\circ}$), we show results for two bracketing cases: $i_{\text{w}} = 90^{\circ}$ and $i_{\text{w}} = 70^{\circ}$. We show representative values of the wind speed for several choices of $\theta$ and $i_{\text{w}}$ in Tab.~\ref{tab:wind_model}.

In principle, the velocity profiles of the cool gas place constraints on the opening angle of the wind cone. \citet{2015ApJ...814...83L} use line splitting in the molecular and neutral phases to infer a relatively narrow opening angle of $\theta \approx 5^{\circ} - 10^{\circ}$. \citet{1998ApJ...493..129S} find a similar opening angle for the warm ionized gas ($\theta \approx 10^{\circ} - 15^{\circ}$), while \cite{1990ApJS...74..833H} infer twice this value ($\theta \approx 30^{\circ}$) for the same phase\footnote{Note that we adopt a different definition of $\theta$ from the one used in these papers, $\theta'$: $\theta = \theta'/2$. We have scaled the numbers cited here appropriately.}. The uncertainty on the inclination angle of the wind cone (and the resulting uncertainty on the wind speed) limits the precision of this measurement. Furthermore, although the cool gas velocity profile suggests that it is primarily found at the edges of the outflow cone, the structural relationship between the cool and hot phases is unknown. Thus, we consider a large range of opening angles for the soft X-ray emitting gas, $\theta = 5^{\circ} - 65^{\circ}$.

In the left panel of Fig.~\ref{fig:wind_cone_model}, we compare the modeled velocity dispersion for $i_{\text{w}} = 90^{\circ}$ with the $90$\% confidence interval measured from \ion{O}{8}. This suggests a fast wind speed ($v_{\text{out}} \gtrsim 2000$ km s$^{-1}$) for all values of $\theta$, with increasingly high values of $v_{\text{out}}$ required for decreasing $\theta$. If the wind cone is moderately inclined with respect to our line of sight ($i_{\text{w}} = 70^{\circ}$; see the right panel of Fig.~\ref{fig:wind_cone_model}), then only slightly lower wind speeds can be accommodated. These wind speeds are well above the escape velocity of M82 ($v_{\text{esc}} \lesssim 450$ km s$^{-1}$; \citealt{2009ApJ...697.2030S}). It is important to note that this simple biconical wind model neglects potentially important aspects of the wind geometry (e.g., the launching of the wind from a circumnuclear ring of star clusters and the contribution of shocked ISM to the wind emission). The wind speeds implied by the model should therefore be considered as illustrative rather than quantitatively predictive.

We can compare the inferred wind speed of the $T \approx 3 \times 10^{6}$ K gas to that of the hotter, hard X-ray emitting gas and the warm ionized and neutral phases. Based on the temperature inferred from the hard X-ray spectrum\footnote{A diffuse, thermal component is seen in \ion{Fe}{25} 6.7 keV emission in the nucleus of M82 \citep[e.g.,][]{2014MNRAS.437L..76L}. While discrete sources, including supernova remnants, may contribute to the \ion{Fe}{25} emission, the diffuse component likely accounts for at least $\approx 50$\% of the emission \citep{2021A&A...652A..18I}.}, \citet{2009ApJ...697.2030S} find a terminal wind speed of $v_{\text{out}} \approx 1400 - 2200$ km s$^{-1}$. In contrast, the inferred outflow speed for the $T \lesssim 10^{4}$ K gas is $v_{\text{out}} \approx 500 - 600$ km s$^{-1}$ from \ion{H}{1} \citep{2018ApJ...856...61M} and $v_{\text{out}} \approx 525 - 655$ km s$^{-1}$ from H$\alpha$ \citep{1998ApJ...493..129S}. Our simple model thus provides reasonable solutions for the soft X-ray wind speed (e.g., $v_{\text{out}} \approx 2000$ km s$^{-1}$, $\theta \approx 60^{\circ}$) that fall between the literature values for the colder phases and the upper bound for the hotter gas. The energy budget of the $T \approx 3 \times 10^{6}$ K gas is dominated by the kinetic energy of the outflow, with an energy-loading factor (ratio of kinetic to thermal energy) of $E_{\text{kin}}/E_{\text{th}} \gtrsim 25$ for $v_{\text{out}} \gtrsim 2000$ km s$^{-1}$.

The inferred wind speed is towards the upper end of the range predicted by hot wind models for the $T \gtrsim 10^{6} - 10^{7}$ K gas, for outflows with low to moderate mass-loading factors \citep[e.g.,][]{1985Natur.317...44C, 2016MNRAS.455.1830T, 2022ApJ...924...82F}. For example, a dominant wind temperature of a few $\times 10^{6}$ K and a velocity of $v_{\text{out}} \approx 2000$ km s$^{-1}$ are predicted at a radius of $R = 2$ kpc for a mass-loading factor of $\dot M_{\text{hot}}/\dot M_{\star} \approx 0.2$ and a star-formation rate surface density similar to that of M82 by \citet{2016MNRAS.455.1830T}. This mass loading factor is consistent with the $\dot M_{\text{hot}}/\dot M_{\star}$ = 0.2 - 0.5 found by \citet{2009ApJ...697.2030S} for the M82 outflow. This broad consistency suggests that the \ion{O}{8} ions may indeed arise from the primary outflow rather than from interfaces or shocked ambient material. Following \citet{2009ApJ...697.2030S}, a terminal wind speed of $v_{\text{out}} \approx 2000$ km s$^{-1}$ implies a high central temperature of $kT \approx 12$ keV, or $T \approx 10^{8}$ K. Turbulent motions in the gas may also contribute to broadening the \ion{O}{8} lines, somewhat reducing the implied wind speed and central temperature. However, the role of turbulence is likely modest at best. With a sound speed of $\sigma_{\text{th}} \sim 200$ km s$^{-1}$, the turbulence in the \ion{O}{8}-bearing gas would be highly supersonic if it accounts for the majority of the line broadening, a scenario that would result in rapid, shock-induced cooling and is not energetically favored.

In Fig.~\ref{fig:wind_cone_model}, we also show the $95$\% upper limit on the velocity dispersion derived from \ion{Mg}{12} for the $T \approx 10^{7}$ K gas. The implied wind speed is lower than for the \ion{O}{8}-bearing gas ($v_{\text{out}} \lesssim 1500$ km s$^{-1}$, unless $\theta \lesssim 20^{\circ}$ and $i_{\text{w}} \approx 90^{\circ}$), and the inferred opening angle is smaller at a fixed value of $v_{\text{out}}$. This may be evidence of a change in the wind properties as it moves from the inner $\sim 1$ kpc traced by \ion{Mg}{12} to the several kpc scales traced by \ion{O}{8}. The wind speed rises with height above the disk for an adiabatically expanding, thermal-pressure driven wind \citep{1985Natur.317...44C}, and transonic solutions with monotonically increasing $v_{\text{out}}$ can remain when gravity and radiative losses are included \citep[e.g.,][]{2016ApJ...819...29B}. Additionally, $\theta$ is predicted to increase with height above the disk within several kpc of the midplane by some hydrodynamic models as a wind that is launched perpendicularly to the disk breaks out of the ISM \citep[e.g.,][]{2022arXiv221203898S}. In summary, it is physically plausible that a rising $v_{\text{out}}$ or $\theta$ with distance from the disk can explain the different velocity dispersions observed for the $T \approx 3 \times 10^{6}$ K and $T \approx 10^{7}$ K gas. Alternatively, the \ion{O}{8} and \ion{Mg}{12} ions may have different primary origins; for example, the compact spatial distribution of \ion{Mg}{12} suggests that it may have a significant contribution from shock-heated ISM surrounding the circumnuclear star clusters.

\section{Conclusions}\label{sec:conc}

Using deep archival observations from the Reflection Grating Spectrometer on $XMM-Newton$, we found evidence for a fast soft X-ray wind from the prototypical dwarf starburst galaxy M82. After accounting for the effects of the complex RGS LSF, we measured a broad line-of-sight velocity dispersion ($\sigma_{\text{v}} = 1160^{+100}_{-90}$ km s$^{-1}$) at \ion{O}{8}, implying wind speeds of $v_{\text{out}} \gtrsim 2000$ km s$^{-1}$ for the $T \approx 3 \times 10^{6}$ K gas. Notably, the hotter ($T \approx 10^{7}$ K) and more spatially compact gas traced by \ion{Mg}{12} has a significantly smaller velocity dispersion ($\sigma_{\text{v}} \lesssim 370$ km s$^{-1}$), potentially indicating an increasing wind speed or opening angle with distance from the disk or a primary origin of the \ion{Mg}{12} emission in shock-heated ISM surrounding the circumnuclear star clusters. Based on our kinematic analysis, we draw three main conclusions about the origin and emission mechanism of the soft X-ray wind in M82:
\begin{enumerate}[label=(\roman*)]
    \item{\textit{The \ion{O}{8}-bearing gas is likely not found} primarily at interfaces with cool gas. Although there is evidence for CX emission from M82 \citep[e.g.,][]{2014ApJ...794...61Z}, suggesting that interfaces are common, the much broader velocity dispersion at \ion{O}{8} ($\sigma_{\text{v}} = 1160^{+100}_{-90}$ km s$^{-1}$) compared to that at \ion{H}{1} and H$\alpha$ ($\sigma_{\text{v}} \approx 100 - 300$ km s$^{-1}$) suggests that the $T \approx 3 \times 10^{6}$ K gas is not primarily found at interfaces. RGS provides only an upper limit on the velocity dispersion of the $T \approx 10^{7}$ K gas traced by \ion{Mg}{12} ($\sigma_{\text{v}} \lesssim 370$ km s$^{-1}$), and thus an interface scenario cannot be ruled out for this phase.} 
    \item{\textit{The soft X-ray wind may be too fast to be predominantly produced by shocks induced by a hotter, faster wind.} A possible origin for the soft X-ray emission is a sheath of shocked gas that forms at the edges of the wind cone as a hotter ($T > 10^{7}$ K), faster outflow impacts the ambient medium. Additionally, the emission may arise from shocked gas within the outflow cone, as slower, cooler clouds are overrun by a hot wind. However, if the soft X-ray wind speed is as fast as implied by the \ion{O}{8} velocity dispersion, $v_{\text{out}} \gtrsim 2000$ km s$^{-1}$, this scenario requires a wind speed for the $T > 10^{7}$ K gas that is significantly larger than the terminal velocity of $v_{\text{out}} \approx 1400 - 2200$ km s$^{-1}$ inferred by \citet{2009ApJ...697.2030S} from the temperature of the thermal, diffuse gas seen in \ion{Fe}{25} emission \citep[e.g.,][]{2014MNRAS.437L..76L}. This is thus a less likely scenario for the origin of the soft X-ray emission.}
    \item{\textit{The soft X-ray wind may be produced by cooling of the dominant phase of the outflow.} A wind speed of $v_{\text{out}} \approx 2000$ km s$^{-1}$ for the $T = 3 \times 10^{6}$ K gas is within the range generally predicted by hot wind models, albeit towards the upper end and requiring low mass-loading factors. The relatively low mass-loading factor inferred for the M82 wind by \citet{2009ApJ...697.2030S} ($\dot M_{\text{hot}}/\dot M_{\star}$ = 0.2 - 0.5) is compatible with fast ($v_{\text{out}} \sim 1000 - 2000$ km s$^{-1}$) outflows of a few $\times 10^{6}$ K gas \citep[e.g.,][]{2016MNRAS.455.1830T}. In this scenario, the soft X-ray photons trace the dominant phase of the wind on $\gtrsim 1$ kpc scales, which may have a significant volume filling factor.}
\end{enumerate}

The launch of \textit{XRISM} \citep{2020arXiv200304962X, 2020SPIE11444E..22T} and its Resolve microcalorimeter has opened a new era in high energy resolution X-ray spectroscopy. Performance Verification observations of M82 with \textit{XRISM}-Resolve, obtained with the gate valve closed, will probe hot gas in and around the wind-launching zone via ions including \ion{Si}{14} (2 keV), \ion{S}{15} (2.4 keV), and \ion{Fe}{25} (6.7 keV). These will provide diagnostics of the wind structure and kinematics at the base of the outflow and the total energy carried by the hot wind fluid. A joint synthesis of the \textit{XRISM}-Resolve and \textit{XMM}-RGS observations will thus provide insight into the thermal and structural evolution of the wind from the galactic nucleus to kpc scales.

\begin{acknowledgments}
We thank the anonymous referee for constructive comments and Anna Ogorza\l{}ek for assistance with the RGS data reduction and analysis. This material is based upon work supported by NASA under award number 80GSFC21M0002. This research has made use of NASA’s Astrophysics Data System.
\end{acknowledgments}

%

\vspace{5mm}
\facilities{XMM(RGS), Chandra(ACIS)}




\appendix

\section{On a possible ULX Outflow in Obs 3}
\label{sec:ulx}

\begin{figure}
\centering
\includegraphics[scale = 0.22]{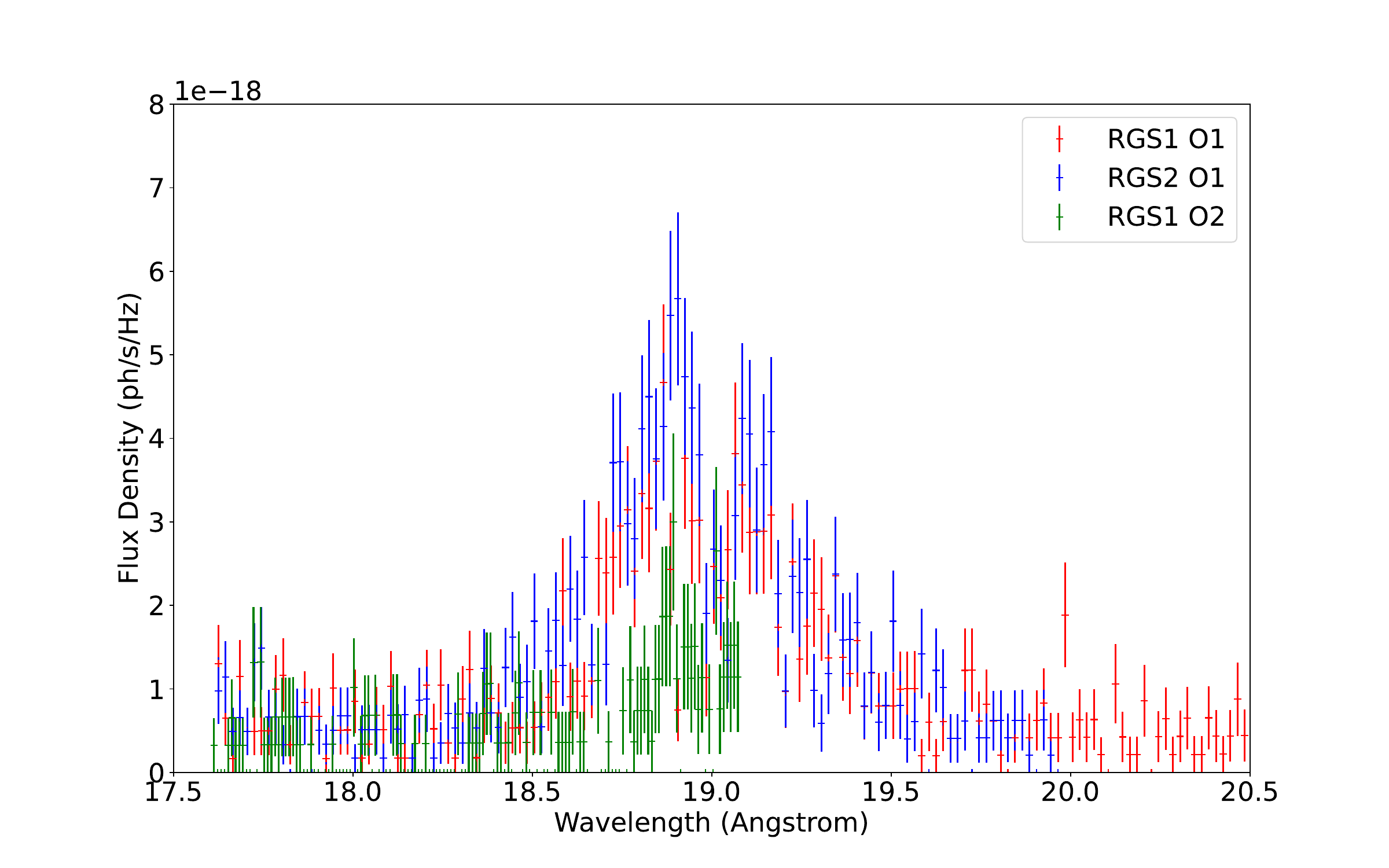}
\includegraphics[scale = 0.22]{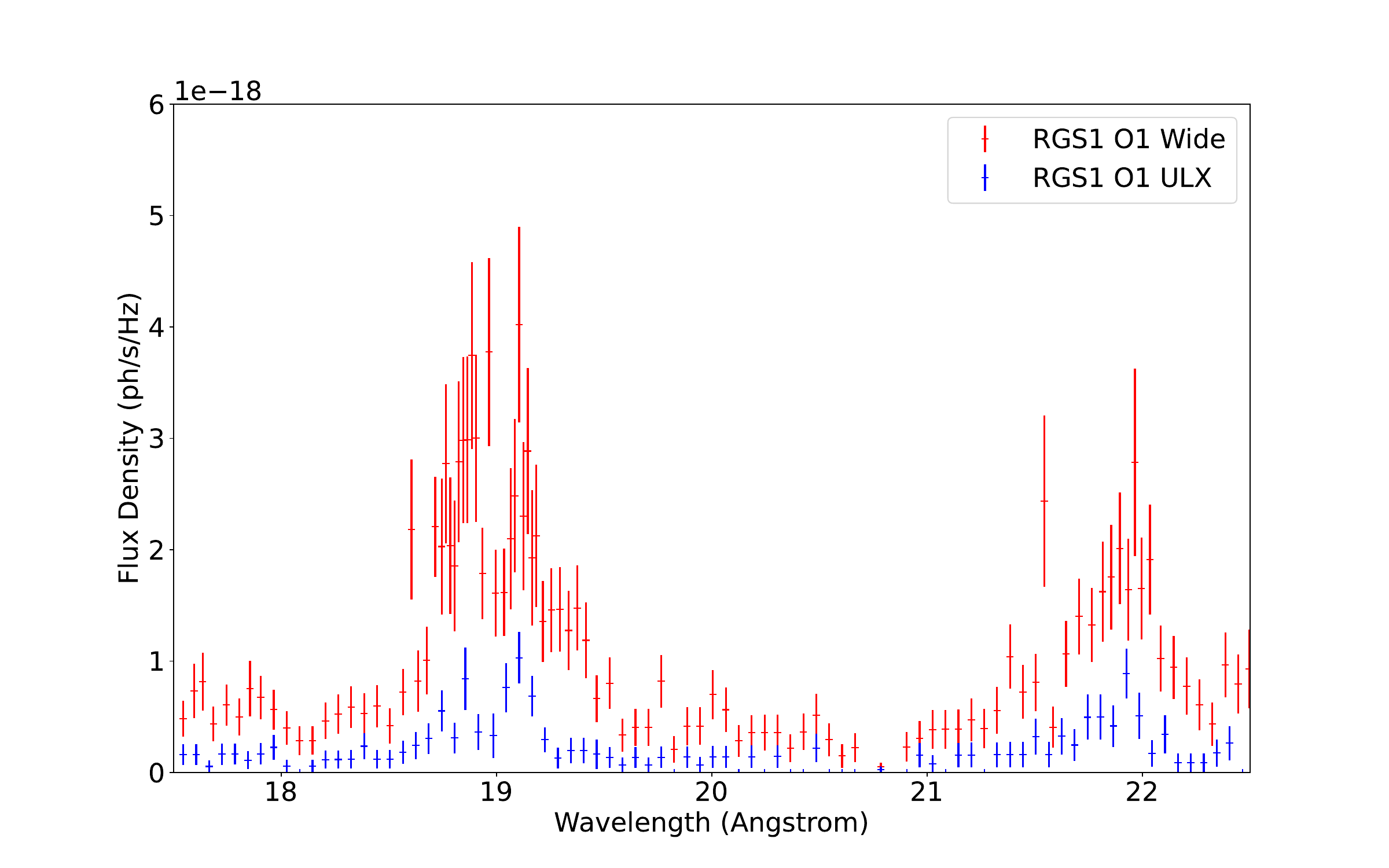}
\caption{Left: The \ion{O}{8} 0.65 keV line in Obs 3 has a double peak at 18.9~\AA\ and 19.15~\AA\ not seen in other observations. The peak is present in both first-order RGS1 (red) and RGS2 (blue) spectra (extracted from the same aperture as described in Section~\ref{sec:data}), and there is tentative evidence for it in the second-order RGS1 spectrum (green, which meets the missing chip in the middle of the line; the \ion{O}{8} line is entirely lost in RGS2). There is no evidence that the line structure results from instrumental artifacts or the surface brightness map that distorts the LSF. Right: Most of the flux in the 19.15~\AA\ peak of the RGS1 spectrum (red) is localized in a $15''$-wide strip in cross-dispersion space coincident with the continuum emission from M82 X-2 (blue). There is a corresponding peak in the anomalous \ion{O}{7} 0.57 keV line. The ``ULX'' spectrum also contains a significant contribution from the galactic wind. Both spectra have been binned to SNR $= 3$ for visual clarity and the corresponding RGS2 spectra have been omitted because they do not extend to \ion{O}{7}.}
\label{fig:obs3_line}
\end{figure}

As noted in Section~\ref{sec:results_OVIII}, the best-fit profile to the \ion{O}{8} 0.65 keV line in Obs 3 has a significantly higher velocity dispersion than in other observations. The line profile (Figure~\ref{fig:obs3_line}) also has a striking double peak at $\lambda \approx 18.90$~\AA\ and 19.15~\AA. Here we report on the evidence that this structure is real, astrophysical, and may have a component associated with one of the ULXs in M82.

\begin{figure}
\centering
\includegraphics[scale = 0.26]{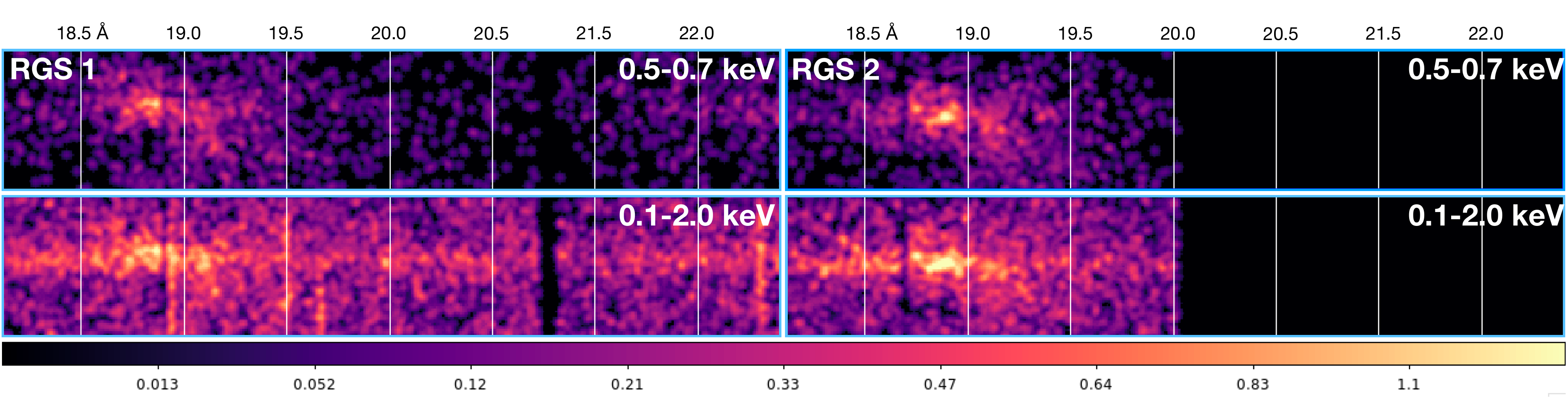}
\caption{Projections of the events in the RGS1 (left) and RGS2 (right) modules in the $\lambda$--cross-dispersion angle plane. The inactive RGS2 chip leaves a black gap at the right. The colorbar and scale are the same in each image, and each image has been smoothed with a Gaussian kernel with $\sigma = 3$~pixels to enhance faint structure. Each panel shows events between 18-22~\AA\ in the dispersion direction and between $-150''$ and $+150''$ in the cross-dispersion direction. The top panels show events with detected energies between 0.5-0.7~keV. The oxygen ``image'' of the M82 nebula is clearly visible. The bottom panels show events between 0.1-2~keV and reveal bad columns (dark or bright columns). No bad columns or other artifacts can explain the double-peaked structure of the Obs 3 \ion{O}{8} 0.65 keV line. The ULX spectrum appears as a narrow horizontal band near zero cross-dispersion in the bottom panels.}
\label{fig:obs3_PI}
\end{figure}

The double-peaked structure cannot be explained by instrumental effects or artifacts. This structure is clear in both RGS modules, and there is a hint of it in the second-order RGS1 spectrum (see Figure~\ref{fig:obs3_line}). This agreement strongly suggests that the peaky structure cannot be explained by bad columns or hot or cool pixels, which differ between modules. We confirmed this by examining the RGS event list. The RGS assigns wavelengths along the dispersion axis based on the pointing and dispersion angle, but orders are sorted and background counts filtered out by filtering events on the energy at which each is detected by the CCDs. While the CCD energy resolution is far worse than that of the RGS, it is sufficient to reject the majority of background counts. Bad columns and pixels often produce false events with recorded energies across the spectrum, so to conduct the most sensitive search for artifacts not flagged by the pipeline, we projected events with pulse invariant (PI) energies between 0.1 and 2 keV onto the $\lambda$--cross-dispersion angle plane (Figure~\ref{fig:obs3_PI}). The only artifacts found around 19~\AA\ are already included in the RMF produced by the RGS pipeline.

We can also rule out a time-variable component. The 0.1-2~keV light curve includes a background flare in the last few ks of the exposure, but excluding this period does not significantly alter the spectrum, and the double peak remains clearly visible. There are no anomalous periods of spacecraft attitude during the observation, nor are other isolated emission lines distorted, so we can likewise rule out problems with pointing or attitude knowledge. The double-peaked shape is consistent between the first and second halves of the observation, and there are no significant features in the light curve beyond the flare at the end.

\begin{figure}
\centering
\includegraphics[scale = 0.26]{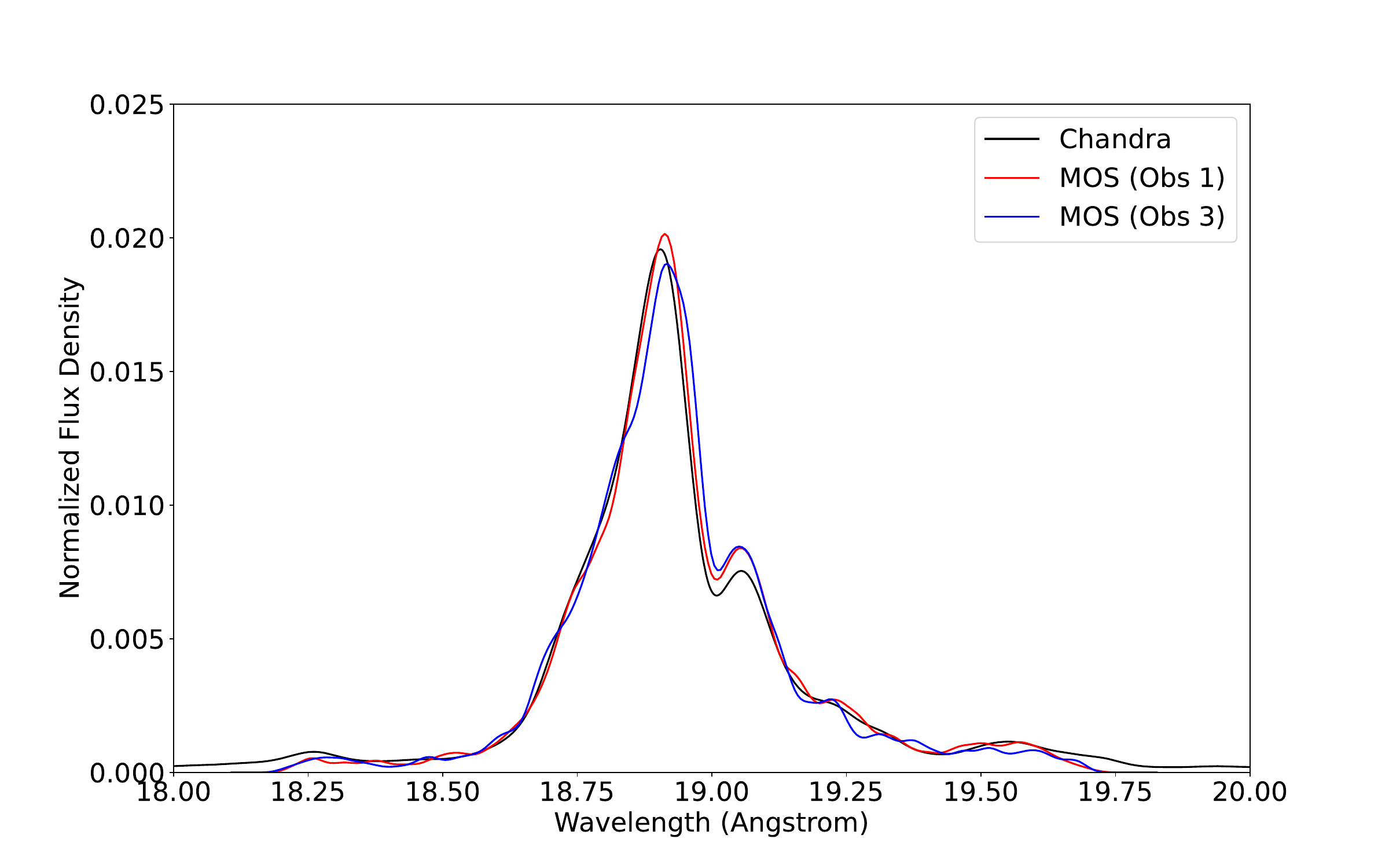}
\caption{LSFs created for the Obs 3 spectrum at \ion{O}{8} 0.65 keV, with a position angle of 138$^{\circ}$ and an extraction aperture of $1'$. The LSF created from \textit{Chandra} data is shown in black, while LSFs created from the Obs 1 and Obs 3 MOS images are shown in red and blue, respectively. The similarity between the LSFs demonstrates that the double-peaked feature in the Obs 3 \ion{O}{8} line cannot be explained by a feature present in the Obs 3 image and also that one need not construct an LSF for each observation from the MOS images for that observation.}
\label{fig:obs3_lsf}
\end{figure}

The double-peaked structure cannot result from the distortion of the LSF produced by the surface brightness distribution of the M82 nebula. We created LSFs for the surface brightness at 0.65 keV using the \textit{Chandra} image, the MOS images from Obs 3, and the MOS images from Obs 1, appropriately rotated based on the 138$^{\circ}$ PA of Obs 3. The profiles are in remarkable agreement with each other but not the actual Obs 3 spectrum (Figure~\ref{fig:obs3_lsf}). There is a secondary peak on the shoulder of the main peak of the LSF, but it is too weak to produce the structure seen in Obs 3.

If the double peak is real and unrelated to structure in the surface brightness map, but is only found in Obs 3, the most likely source is a transient or variable source. Obs 3 targeted M82 X-2 and caught it in a high flux state in which the ULX continuum was stronger relative to the thermal emission from the galactic wind nebula than in the other observations. We thus investigated whether the ULX can explain the double peak by extracting a spectrum from a $15''$ aperture (in the cross-dispersion direction) at the position of the ULX (Figure~\ref{fig:obs3_line}). This spectrum also includes a significant amount of galactic wind emission but clearly shows that the peak near 19.1~\AA\ comes from the narrower aperture. Apertures excluding this region do not have a second peak. The association of the 19.1~\AA\ peak with the ULX suggests that the \ion{O}{8} line includes a narrow component from the ULX, but with a large redshift ($z \approx 0.01c$). Mildly relativistic, photoionized outflows are seen in emission lines, including \ion{O}{8}, in other ULXs \citep{Pinto2017,Pinto2021}.

There is additional supporting evidence for a ULX association from the \ion{O}{7} 0.57 keV triplet, which differs strikingly between Obs 3 and the other observations in that the peak occurs near 21.9~\AA\ in Obs 3 (Figure~\ref{fig:obs3_line}), whereas the forbidden line is clearly strongest and stretched by the effect of the surface brightness distribution in the other observations. Assuming that 21.9~\AA\ corresponds to a shifted 21.602~\AA\ resonant line, the redshift would be 0.014$c$. No other lines show such differences, although it is worth noting that a 0.01$c$ split between two peaks would be increasingly hard to see at shorter wavelengths because the distortion of the LSF by the surface brightness distribution is independent of wavelength (aside from wavelength-dependent variation in the surface brightness itself) and the lines in the Fe-L complex are highly blended. Likewise, the complexity of the nebular emission makes it impossible to search for absorption lines from the ULX.

ULX flux does not straightforwardly connect to the intensity of any photoionized or collisionally ionized outflow \citep{Pinto2021}, so the absence of similar features in other observations does not cleanly map to the lower ULX flux in those exposures. Nevertheless, if these features trace a ULX wind then the wind must be transient, as the other observations would be sensitive to secondary peaks at the level found in Obs 3.

Finally, we note that adding a narrow Gaussian line at the wavelength of the 19.1~\AA\ peak does not bring the velocity dispersion of the primary component into agreement with the other observations. The red shoulder of the line (Figure~\ref{fig:obs3_line}) is broader than expected from the LSF (Figure~\ref{fig:obs3_lsf}), so $\sigma > 1500$~km~s$^{-1}$ is still preferred.

To summarize, the double-peaked \ion{O}{8} line in Obs. 3 appears to be real and most of the photons at 19.1~\AA\ are associated with a narrow aperture containing most of the ULX continuum. This tentative association would imply a 0.01$c$, redshifted outflow seen in emission but cannot be definitively proven from Obs 3 alone. Nevertheless, there is sufficient evidence to exclude Obs 3 from the \ion{O}{8} measurement of velocity broadening in the galactic wind nebula.



\bibliographystyle{aasjournal}



\end{document}